\documentclass[sigconf,screen,urlbreakonhyphens=false]{acmart}

\acmSubmissionID{fsews24aiwaremain-p52-p}

\copyrightyear{2024}
\acmYear{2024}
\setcopyright{rightsretained}
\acmConference[AIware '24]{Proceedings of the 1st ACM International Conference on AI-Powered Software}{July 15--16, 2024}{Porto de Galinhas, Brazil}
\acmBooktitle{Proceedings of the 1st ACM International Conference on AI-Powered Software (AIware '24), July 15--16, 2024, Porto de Galinhas, Brazil}
\acmDOI{10.1145/3664646.3665664}
\acmISBN{979-8-4007-0685-1/24/07}

\received{2024-04-05}
\received[accepted]{2024-05-04}

\ccsdesc[500]{Software and its engineering~Software verification and validation}

\usepackage{booktabs}
\usepackage{xspace}
\usepackage{ifthen}
\usepackage{listings}
\usepackage[noabbrev,nameinlink]{cleveref}
\usepackage{microtype}\newboolean{draft}

\setboolean{draft}{false}

\ifthenelse{\boolean{draft}}
  {
    \newcommand{\todo}[1]{{\color{red} \textbf{todo: #1}\xspace}}
  }
  {
    \newcommand{\todo}[1]{\relax}
  }
  
\def\toolname{AutoCommenter\xspace}
\def\companyname{Google\xspace}

\def\bpFormatting{Formatting\xspace}

\def\bpNaming{Naming\xspace}

\def\bpDocumentation{Documentation\xspace}

\def\bpLanguageFeatures{Language features\xspace}

\def\bpCodeIdioms{Code idioms\xspace}

\def\GoogleZRH{Google}
\def\ZRH{Zurich}
\def\CH{Switzerland}
\def\GoogleUS{Google}
\def\US{USA}
\def\MTV{Mountain View}
\def\SVL{Sunnyvale}
\def\SEA{Seattle}
\def\MTR{Montreal}
\def\GoogleCA{Google}
\def\CA{Canada}
\def\UW{University of Washington}

\author{\mbox{Manushree Vijayvergiya}}
\orcid{0009-0006-3957-5399}
\email{manushree@google.com}
\affiliation{%
  \institution{\GoogleZRH}
  \city{\ZRH}
  \country{\CH}
}

\author{Małgorzata Salawa}
\orcid{0009-0003-4904-7109}
\email{magorzata@google.com}
\affiliation{%
  \institution{\GoogleZRH}
  \city{\ZRH}
  \country{\CH}
}

\author{Ivan Budiselić}
\orcid{0009-0008-6696-3842}
\email{ibudiselic@google.com}
\affiliation{%
  \institution{\GoogleZRH}
  \city{\ZRH}
  \country{\CH}
}

\author{Dan Zheng}
\orcid{0009-0006-4523-6262}
\email{danielzheng@google.com}
\affiliation{%
  \institution{\GoogleUS}
  \city{\MTV}
  \country{\US}
}

\author{Pascal Lamblin}
\orcid{0009-0009-9746-2001}
\email{lamblinp@google.com}
\affiliation{%
  \institution{\GoogleCA}
  \city{\MTR}
  \country{\CA}
}

\author{Marko Ivanković}
\orcid{0000-0002-8548-6008}
\email{markoi@google.com}
\affiliation{%
  \institution{\GoogleZRH}
  \city{\ZRH}
  \country{\CH}
}

\author{Juanjo Carin}
\orcid{0009-0002-6099-3940}
\email{juanjocarin@google.com}
\affiliation{%
  \institution{\GoogleUS}
  \city{\SVL}
  \country{\US}
}

\author{Mateusz Lewko}
\orcid{0009-0007-5192-7913}
\email{mlewko@google.com}
\affiliation{%
  \institution{\GoogleZRH}
  \city{\ZRH}
  \country{\CH}
}

\author{Jovan Andonov}
\orcid{0009-0006-6823-2790}
\email{jandonov@google.com}
\affiliation{%
  \institution{\GoogleZRH}
  \city{\ZRH}
  \country{\CH}
}

\author{Goran Petrović}
\orcid{0000-0002-8056-7431}
\email{goranpetrovic@google.com}
\affiliation{%
  \institution{\GoogleZRH}
  \city{\ZRH}
  \country{\CH}
}

\author{Daniel Tarlow}
\orcid{0009-0009-4304-6395}
\email{dtarlow@google.com}
\affiliation{%
  \institution{\GoogleCA}
  \city{\MTR}
  \country{\CA}
}

\author{Petros Maniatis}
\orcid{0000-0003-3777-5291}
\email{maniatis@google.com}
\affiliation{%
  \institution{\GoogleUS}
  \city{\MTV}
  \country{\US}
}

\author{René Just}
\authornote{Work done at Google.}
\orcid{0000-0002-5982-275X}
\email{rjust@cs.washington.edu}
\affiliation{%
  \institution{\UW}
  \city{\SEA}
  \country{\US}
}
 
\keywords{Artificial Intelligence, Code Review, Coding Best Practices}

\title[AI-Assisted Assessment of Coding Practices in Modern Code Review]{AI-Assisted Assessment of Coding Practices\\in Modern Code Review}

\renewcommand{\shortauthors}{Manushree Vijayvergiya et al.}

\begin{abstract}
Modern code review is a process in which an incremental code contribution made by a code author is reviewed by one or more peers before it is committed to the version control system. An important element of modern code review is verifying that code contributions adhere to best practices. While some of these best practices can be automatically verified, verifying others is commonly left to human reviewers.
This paper reports on the development, deployment, and evaluation of \toolname, a system backed by a large language model that automatically learns and enforces coding best practices. We implemented \toolname for four programming languages (C++, Java, Python, and Go) and evaluated its performance and adoption in a large industrial setting. Our evaluation shows that an end-to-end system for learning and enforcing coding best practices is feasible and has a positive impact on the developer workflow. Additionally, this paper reports on the challenges associated with deploying such a system to tens of thousands of developers and the corresponding lessons learned.
\end{abstract}

\begin{document}
\maketitle

\section{Introduction}
\label{sec:intro}

Modern code review~\citep{moderncodereview,10.1145/2491411.2491444} (compared to holistic code review~\citep{5388086}) has grown organically over the years in open-source and industrial settings. A set of common peer-review criteria have emerged~\cite{10.1145/2491411.2491444,6148202,6606617}, which include coding best practices. Many companies, projects, and even programming languages formally define them in the form of ``style guides''~\cite{googlestyleguides,linuxstyleguide,pythonstyleguide,ruststyleguide} that commonly cover the following aspects:

\begin{itemize}
\item \textit{\bpFormatting}: line limits, use of whitespaces and indentation, placement of parentheses and brackets, etc.;
\item \textit{\bpNaming}: capitalization, brevity, descriptiveness,  etc.;
\item \textit{\bpDocumentation}: expected placement and content of file-level, function-level, and other comments;

\item \textit{\bpLanguageFeatures}: use of specific language features in different (code) contexts;
\item \textit{\bpCodeIdioms}: use of code idioms to improve code clarity, modularity, and maintainability.
\end{itemize}
Developers generally report high satisfaction with modern code review processes~\cite{moderncodereview,winters2020software}.
One of their main benefits is the learning experience for code authors who are not familiar with the codebase, specific language features, or common code idioms. During a review, an expert developer educates the code author on best practices, in addition to reviewing (and learning about) the code contributions and their implications.

Static analysis tools such as linters~\cite{johnson1977lint} can automatically verify that code adheres to some best practices (e.g., formatting rules), and some tools can even automatically fix violations. However,
nuanced guidelines or those with exceptions are difficult to automatically verify in their entirety (e.g., naming conventions and justified deviations in legacy code),
and some guidelines cannot be captured by precise rules at all (e.g., clarity and specificity of code comments) and rely on human judgement and collective developer knowledge.
As a result, it is generally expected that human reviewers check code changes for best practice violations.

The biggest cost of the code-review process is the time required, especially from expert developers. Even with significant automation in place, and keeping the process as lightweight as possible, a developer can easily dedicate several hours daily to this task~\cite{moderncodereview}.

Recent advances in machine learning, capabilities of large language models (LLMs) in particular, suggest that LLMs are suitable for code-review automation (e.g.,~\cite{tufano2024code,tufano2022using,hong2022commentfinder,li2022auger,10.1145/3540250.3549081,thongtanunam2022autotransform}). However, the software engineering challenges around deploying an end-to-end system at scale remain unexplored. Likewise, extrinsic evaluations of such systems on overall efficacy and user acceptance are missing.

This paper investigates whether it is possible to partially automate the code-review process, specifically the detection of best practice violations, thereby providing timely feedback for code authors and allowing reviewers to focus on overall functionality.
Specifically, this paper reports on our experience of developing, deploying, and evaluating \toolname---an automated code-review assistant---in an industrial setting at \companyname, where it is currently used by tens of thousands of developers every day.

\smallskip\noindent
In summary, the contributions of this paper are: 

\begin{itemize}
\item A general architecture of an LLM-based code-review assistant system (\cref{sec:approach}).
\item A description of tool calibration and deployment to tens of thousands of developers (\cref{sec:deployment}).
\item An evaluation of the system (\cref{sec:eval}).
\item A summary and discussion of lessons learned (\cref{sec:lessons}).
\end{itemize}\section{Background}
\label{sec:background}

\toolname was developed in a large industrial setting at \companyname. The modern code review practices at \companyname are similar to those of other industrial and open source projects~\cite{moderncodereview}.

\subsection{Code Review Process}
\label{sec:background:code_review}

The code review process at \companyname is well established, change-based, and tool-assisted.
\citet{IvankovicPJF2019} and \citet{petrovic2023resolution} provide a detailed summary of the process.
Each change to the codebase must be reviewed by at least one other developer. Every day, tens of thousands of changes to the codebase go through the review process and tens of thousands of developers participate in the process, as both code authors and reviewers.

Authors and reviewers exchange comments through the code review system, and a review progresses through snapshots of files affected by the change. Each reviewer comment is attached to a specific line and column range in a specific file snapshot. To resolve a comment, the author typically modifies the file in their local copy and exports a new snapshot for the next round of code review. When the author and all reviewers are satisfied and no automated analysis is blocking the merge, the code is merged into the codebase.

The most expensive part of the code review process is the time spent by code authors and reviewers ``shepherding'' a change (from initial coding, through addressing reviewer comments and ensuring all automated analyses pass, to finally merging the change into the codebase). While the process is optimized with automated systems analyzing the code before the review (notably automatic code formatting without human intervention), code reviews still cost thousands of developer-years per year. Thus, even single-digit percentage savings translate into significant business impact.

\subsection{Best Practices}
\label{sec:background:best_practices}
\begin{figure}
  \centering
  \frame{\includegraphics[width=.98\linewidth]{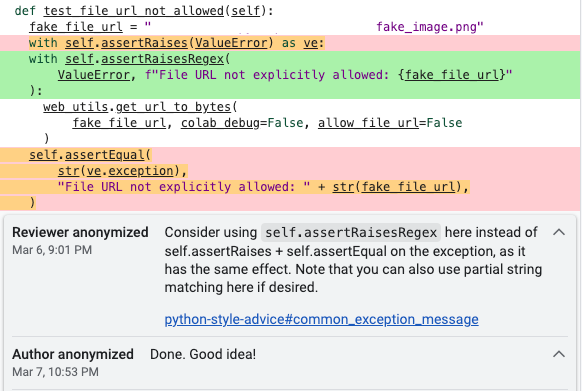}}
  \vspace*{-6pt}
  \caption{Example comment posted by a human reviewer.
  \label{fig:readability_comment}}
  \vspace*{-6pt}
\end{figure}

A \textit{best practice} is a specific use of programming language that is considered superior, and
a \textit{best practice document} describes how it should be applied and what benefits it brings.
\textit{Best practice URL} refers to a best practice document or specific section therein, and
\textit{best practice violation} refers to a specific piece of code that does not adhere to a best practice, but can be changed to do so.
If clear from context, we use the terms \emph{URL} and \emph{violation} to refer to best practice URL and best practice violation, respectively.

\companyname's central code repository contains code in many languages, with C++, Java, Python and Go exceeding 100 million lines each~\cite{googlemonorepo2016}. For 15 different languages there are formal style guides readily available to all developers. Many of these languages have additional language primers, documentation for core libraries, and tip-of-the-week style newsletters. While these materials are not as strictly enforced as style guides, they are frequently referenced in code reviews. Some languages boast hundreds of pages of such documentation. Both code authors and reviewers are expected to verify that the code follows all best practices.

A formal mechanism called ``readability'', introduced more than a decade ago, ensures that best practices are followed consistently. Dedicated style experts in a given language, called ``readability mentors'', guide inexperienced developers towards proficiency in the language~\cite{moderncodereview}. Readability mentors commonly summarize a best practice in a few sentences and at the end of the comment include a URL for the change author as a reference. Figure~\ref{fig:readability_comment} shows an example of a comment posted by a readability mentor.

\begin{figure*}
  \centering
  \includegraphics[width=.95\linewidth]{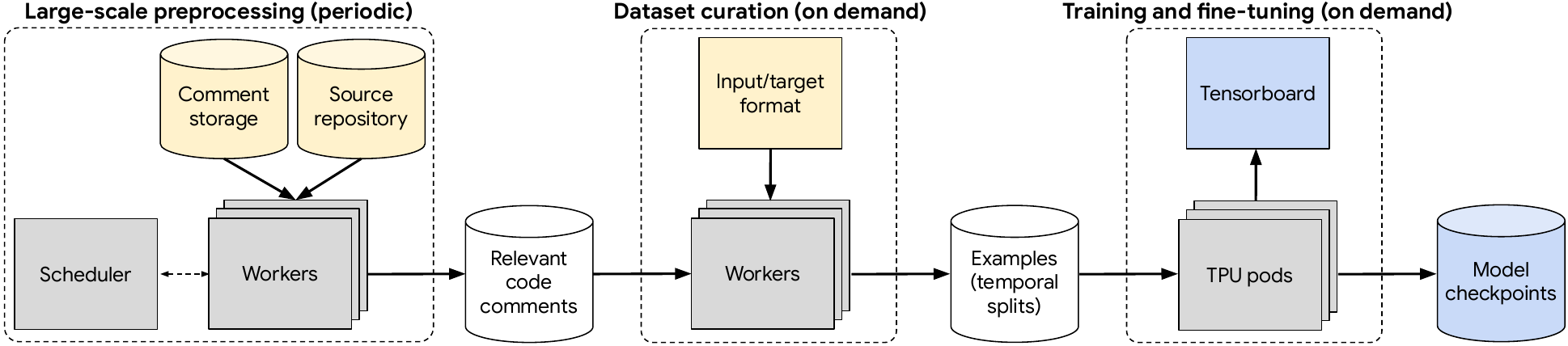}
  \vspace*{-4pt}
  \caption{Architecture of the model-training pipeline.
  \label{fig:training_architecture}}
\end{figure*}
The readability process has some drawbacks. For authors, it increases development time due to additional review rounds. For readability mentors, it can become a monotonous and time-consuming task. It requires mastering hundreds of evolving best practices, including the identification and deprecation of outdated rules, and documenting them (with relevant links) in the code-review system. Additionally, it requires tracking, sometimes through multiple iterations, ensuring that all violations have been remedied.\section{Approach}
\label{sec:approach}

In response to the challenges described~\cref{sec:background:best_practices,sec:background:code_review}, we developed \toolname, a code analysis tool that automatically detects best practice violations. It aims to provide timely feedback for code authors and to alleviate the need for manual best-practice reviews, thereby allowing reviewers to focus on code functionality.

\subsection{Model and Task Definition}
\label{sec:model}

Automating best practice analysis requires a model that can represent source code, pinpoint violation locations, and identify the violated best practice.
We target a text-to-text transformation using a traditional transformer approach based on T5, using T5X~\cite{roberts2023scaling}.

The best practice analysis is one task in a multi-task large sequence model. In addition to the standard pretraining task for T5, span denoising (predicting masked tokens), other tasks used to train this model include code-review comment resolution, next edit prediction, variable renaming, and build-error repair~\cite{C2CGoogle2024}. The training corpus consists of over 3 billion examples, of which the best practice analysis dataset contributes about 800k examples.
The model was trained using the standard cross-entropy loss, typical for such models, and tuned to maximize the sequence accuracy metric, predicting the exact target text for each example.

For the best practice analysis, the \textbf{input} to the model is a task prompt and source code, and the \textbf{target} is a source code location and a URL for a best practice violation.
The task prompt is formatted as a fixed-text code comment, using the programming language's appropriate commenting style. It describes the task in natural language and precedes the source code, which is a direct textual representation of one file. If the input exceeds the model context window, it is truncated.
The location is a byte offset in the source code, and the URL references the violated best practice.
A domain specific language defines the target format, and a special case is the ``empty'' target, if there are no violations.
In addition to the target, \mbox{the model outputs a confidence score ranging from 0 to 1.}

Consider the following input/target example for the Go language.

\begin{center}\textbf{Input}\end{center}
\vspace*{-4pt}
\begin{lstlisting}[language=Go, frame=tb, basicstyle=\small]
// [*] Task: Check language best practices.
// Package addition provides Add
package addition

// Return a sum
func Add(value1, value2 int) int {
	return value1 + value2
}
\end{lstlisting}

\begin{center}\textbf{Target}\end{center}
\vspace*{-4pt}
\begin{lstlisting}[language=Go, frame=tb, basicstyle=\small]
INSERT 153 COMMENT https://go.dev/doc/comment#func
\end{lstlisting}
\smallskip
The first line of the input is the fixed-text task prompt; the rest is the source code. The target gives the location (byte offset 153 corresponds to the start of the \texttt{Add} function) and a \texttt{go.dev} URL, pointing to the exact part of the Go language style guide that the function comment violates (in this case, the usual practice of starting a comment with the function name).
Note that the target may contain no, one, or multiple (concatenated) location-URL pairs, depending on the number of violations in the source code.

\subsection{Model Training}

\Cref{fig:training_architecture} shows the architecture of the model-training pipeline, which consists of three parts.
We split dataset creation into two steps (preprocessing and curation) because the first step is significantly more expensive as it operates on a much larger amount of data. The output of the preprocessing step is agnostic to the model's input/target representation.
This separation improves feature velocity by enabling quick iterations on example representations and other example-level adjustments.
The preprocessing step uses a fault-tolerant scheduling system and periodically extracts relevant code comments to ensure that new data is readily available.

\subsubsection{Large-scale preprocessing}
The training examples are created from real code review data, but not all code comments are suitable for model training. Therefore, the preprocessing step, identifies \textit{relevant code comments}---human authored comments that contain a URL pointing to a best practice document. For each comment, the preprocessing step then collects the corresponding source code and relevant metadata, including the comment's location in the source code and its creation time. The output of this step is a set of relevant code comments, each with all the data necessary for curating examples for model training.

\subsubsection{Dataset curation}
Dataset curation is a single, on-demand processing step, implemented as a Beam\footnote{\url{https://beam.apache.org/}} pipeline. It converts each relevant code comment, based on the input/target format described in~\cref{sec:model}, into the standard TensorFlow \textit{Example} data structure.

\subsubsection{Training and fine-tuning}
The curated examples are used directly for model training and evaluation. We use the T5X framework~\cite{roberts2023scaling} on a fleet of TPUs, store the model checkpoints every 1000 steps, and use Tensorboard for monitoring the training.

\subsection{Model Selection}
\label{sec:model_selection_infrastructure}
Two intrinsic evaluations on historical data inform our selection of a model checkpoint, confidence thresholds, and a decoding strategy.
First, an evaluation on the validation and test datasets provides estimates of precision and recall on a per-file basis.
Second, an evaluation on full historical code reviews provides an estimate of the total number of comments per code review, indicating how often developers would interact with \toolname.

\subsubsection{Evaluation on Validation and Test Datasets}
\label{sec:evaluation_historical_comments}

We temporally split the dataset to ensure that the model has not been trained on future code-review snapshots of the code comments in the validation and test datasets.
In our dataset, 85\% of files have exactly one relevant code comment, 11\% have two, and 4\% have three or more.
We define a prediction to be \textit{correct} if the predicted code location(s) and URL(s) match the expected values, regardless of order.

Recall that the model provides a confidence score for each prediction, which introduces another parameter: a prediction can be suppressed if its confidence score is below some threshold $t$. We define $Precision_t$ as the number of correct predictions whose confidence score is greater than $t$ divided by the number of all predictions whose confidence score is greater than $t$; we define $Recall_t$ analogously.
These definitions allow us to estimate how many (in)correct results would be shown to a user, as a function of $t$.
$Precision_t$ and $Recall_t$ are used for model checkpoint comparisons during training.

While this evaluation avoids data leakage and allow us to automatically evaluate model performance, it has a limitation: while it is reasonable to assume that the human comments for a given code-review snapshot are correct, they are not exhaustive. In other words, it is possible that the code in a given code-review snapshot could be improved according to multiple best practices, but a human reviewer did not post comments (with URLs) for all of them. This can happen for several reasons:

\begin{itemize}
\item \textit{Missing references}: A reviewer may comment on an issue, but did not include a URL as a reference. 
\item \textit{Selective commenting}: A reviewer may comment on an issue once, expecting the author to apply a fix throughout.
\item \textit{Varied expertise or focus}: A reviewer may not be familiar with all best practices, or simply choose not to comment on an issue in the context of a given code review (e.g., focusing only on changed code).
\end{itemize}

While most files in our dataset have only one relevant comment, anecdotal evidence based on manually inspecting ``incorrect'' predictions suggests that multiple best-practice comments are typically possible due to the reasons stated above. Given that our ground-truth data is incomplete, our precision and recall measures are noisy. Therefore, we employ a complementary evaluation, described next, to increase confidence in overall model performance.

\subsubsection{Evaluation on Full Historical Code Reviews}
\label{sec:evaluation_historical_changelists}

To accurately gauge potential comment volume in a live setting, we evaluate \toolname on a set of historical code reviews, using a specific model checkpoint and threshold. The predicted comments are not retroactively posted in the code review system, but rather logged in a database for analysis. This allows us to estimate the expected posting frequency---both at per-file and per-code-review granularity. Because developers interact with \toolname for an entire set of code changes subject to code review, this evaluation is an important step before production deployment.
As an added benefit, this step allows for further optimizations and assessment of posting frequencies for different user groups, programming languages, etc.

\subsection{Inference Infrastructure}

The core of \toolname is a central best practice analysis service. This service takes as input one or more source files for analysis. For each file, it constructs a model input (\cref{sec:model}), encodes it in the standard TensorFlow \textit{Example} data structure, and queries the model. The model itself is served by a model service that uses TensorFlow's \textit{Example} data structure as a domain agnostic input-output format. Finally, the best practice analysis service performs a series of filtering steps (\cref{sec:deployment}), which suppress low-quality predictions, and returns the remaining predictions.

\subsection{IDE and Code Review Integration}
Developers interact with \toolname's analysis service in two ways---directly through an IDE plugin, or indirectly through the code review system. The code review system is used by all developers at \companyname, and the IDE by almost all of them.

\toolname's comments appear in the IDE as diagnostics marked with a blue curly underline, spanning the relevant code snippet. Hovering over the underlined code reveals the full comment with a concise summary of the best practice, including a clickable link to the relevant best practice document. This embedded information streamlines the workflow for developers by eliminating the need to switch between the IDE and a web browser for unfamiliar best practices. Since comments in the IDE need to be generated in real-time, we aim to generate comments with sub-second latency.

In the code review system, \toolname runs after each update (i.e., on each new code-review snapshot), automatically posting comments if it detects any violations. Comments produced by automated tools are visually similar to comments produced by humans, but have a differently colored background.

\begin{figure}
  \centering
  \frame{\includegraphics[width=\linewidth]{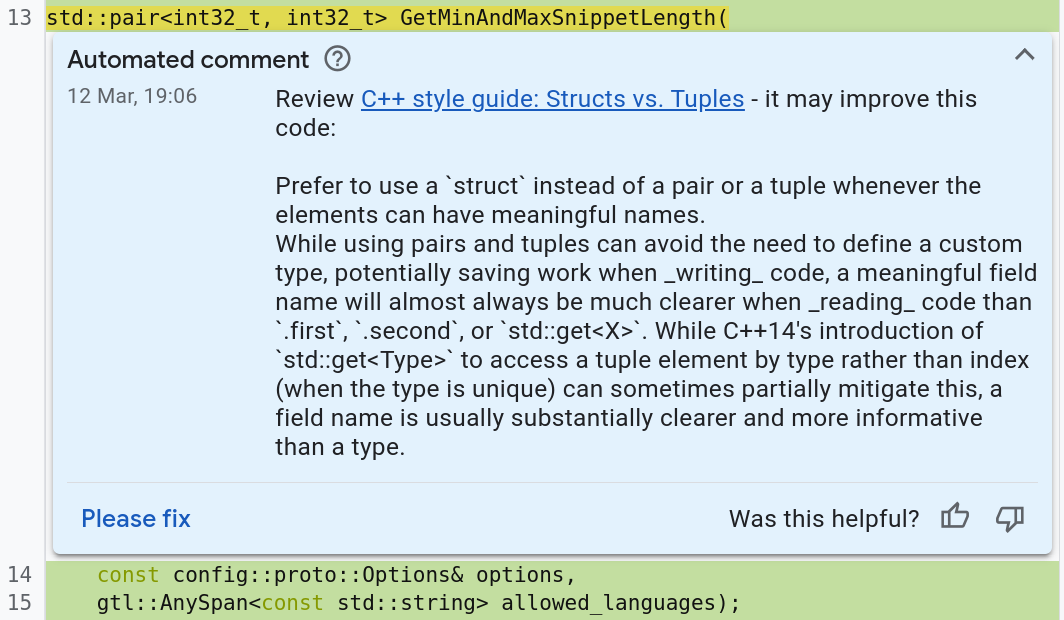}}
  \caption{Example comment posted by \toolname.
  \label{fig:example_critique_comment}}
\end{figure}

\Cref{fig:example_critique_comment} shows an example comment generated and posted by \toolname in the code review system.
Note the thumbs up and thumbs down buttons (right), which authors and reviewers can click if they find a comment particularly useful or not.
Also note the ``Please fix'' button (left), which is visible to reviewers. If clicked, a new comment is generated indicating that the reviewer believes the comment is significant and must be addressed before the code is merged into the codebase.
These feedback buttons are standard in the code review system, present on all comments generated by automated tools (e.g.,~\cite{chen2023murs,prodcov2024}), and provide a signal for a tool's user acceptance. The IDE provides a similar feedback mechanism.
\section{Deployment}
\label{sec:deployment}

We deployed \toolname to all developers at \companyname over a period of time between July 2022 and October 2023:
\begin{itemize}
    \item \textbf{until Jul. 2022---teamfooding}: this paper's authors.
    \item \textbf{Jul. 2022---early adopters}: around 3 thousand volunteers.
    \item \textbf{Jul. 2023---A/B experiment}: about half of all developers.
    \item \textbf{since Oct. 2023---general availability}: all developers.
\end{itemize}
Note that due to industrial confidentiality reasons we are unable to disclose absolute numbers of code reviews, developers, files, comments, or distribution of duration of code reviews. We report on relative measures, where appropriate, and relevant trends.

We continuously evaluated and improved the performance of \toolname, using an iterative refinement approach:

\begin{itemize}
\item Evaluation on historical data (\cref{sec:model_selection_infrastructure}) to get directional insight into how well the model does at the task and to define thresholds and select a decoding strategy.
\item Monitoring and analysis of user interaction and direct feedback through feedback buttons and issue reports.
\item Targeted human evaluation based on patterns observed during other evaluation steps.
\end{itemize}

\begin{figure}
  \centering
  \includegraphics[width=\linewidth]{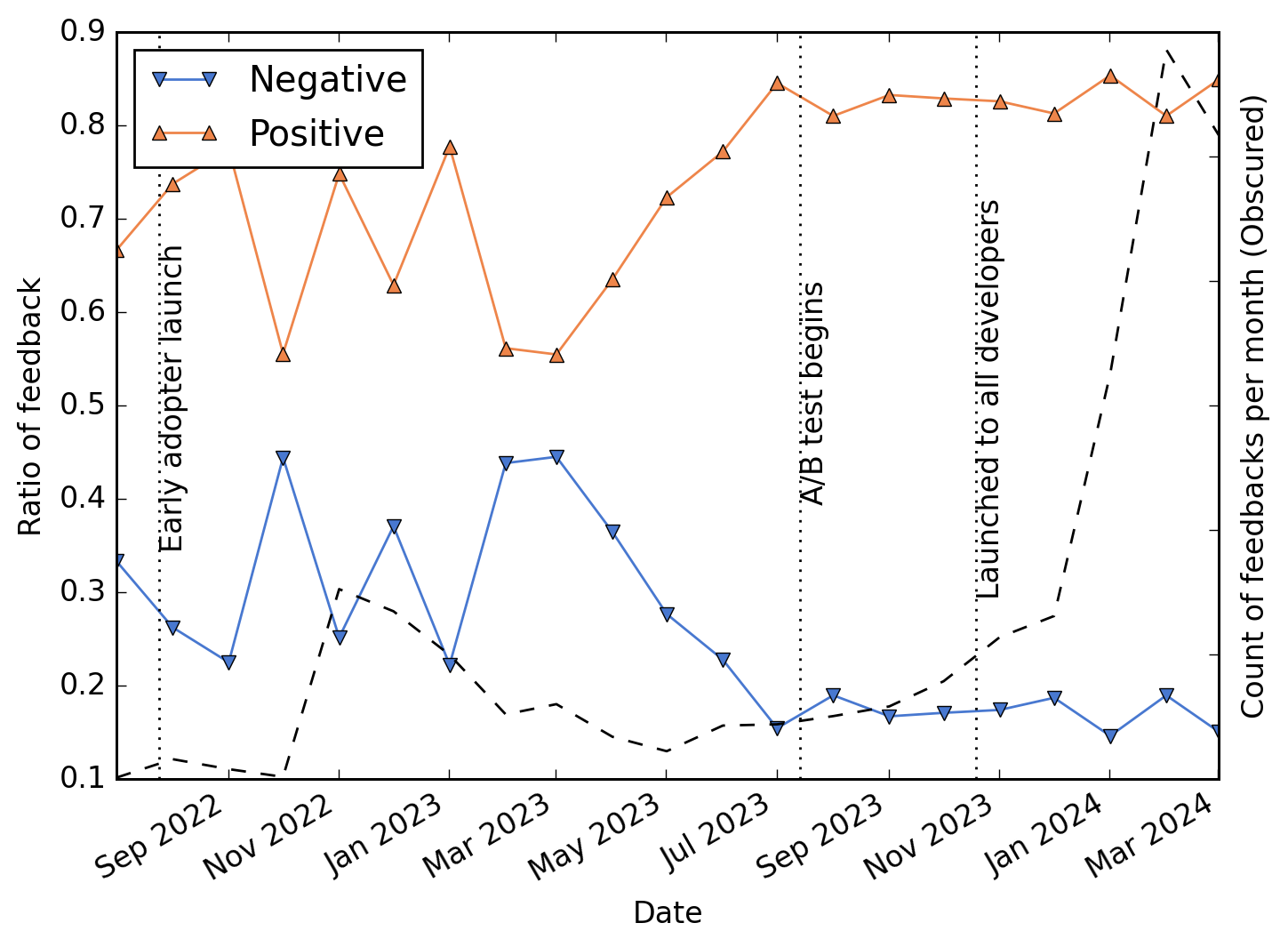}
  \caption{Developer feedback throughout deployment.
  \label{fig:feedback}}
\end{figure}

\Cref{fig:feedback} shows the ratio of positive to negative developer feedback on posted code review comments and IDE diagnostics over time. The dashed line shows the total count of feedback clicks developers provided per month. As is expected, this count is much lower during the early-adopter stage. Additionally, the volatility is higher in this stage because we actively refined \toolname.

Recall the three feedback buttons within the code review system (\cref{fig:example_critique_comment}), which allow developers to express positive and negative sentiment about a posted comment. We consider comments with a thumbs up or ``Please fix'' as positive, and comments with a thumbs down as negative; we define the \textit{useful ratio} as the ratio of positive comments to all comments with feedback.

The remainder of this section describes specific observations and corresponding refinements that we made during deployment.

\subsection{Selecting Threshold and Decoding Strategy}
\label{sec:initial_threshold_and_decoding_strategy_selection}

\subsubsection{Threshold}
During initial deployment we wanted to carefully manage the trust developers had in \toolname and started with a high confidence threshold of $t=0.98$. We manually sampled several hundred results and observed that around 80\% of predictions below the threshold were still correct---that is, the false-negative rate was very high at $t=0.98$.
Additionally, we observed that predictions in Python showed a significantly different distribution of confidence scores, which were disproportionately impacted by thresholding. We conjecture that the training dataset composition (number of distinct URLs and URL frequencies) and specificity of the best practice documents are reasons, but leave a deeper investigation to future work.
An attempt to deploy per-language thresholds proved ineffective as a single threshold per language still did not adequately capture the model's ability to correctly predict hundreds of diverse best practices. This led to a lack of diversity in predicted URLs as the model tended to produce higher scores for some URLs vs. others, irrespective of correctness.
These observations led to the first major change to \toolname: per-URL thresholds computed \mbox{based on the intrinsic evaluation on the validation dataset.}

\subsubsection{Decoding}
An evaluation using per-URL thresholds with greedy decoding on full historical code reviews revealed that \toolname detects violations in 6\% of all changed files. However, 80\% of comments would have been posted on lines of code not modified by the author. Developers typically do not take action on unchanged code. Consequently, \toolname filters generated comments on unchanged lines of code, reducing the ratio of comments in changed files to 1.3\%.
In order to increase this ratio, we experimented with different decoding strategies: greedy (default), beam search, top-k, and top-p sampling. We settled on beam search (generating $n=4$ potential responses), which tripled the posting frequency to 3.9\%. It also yielded a substantially higher URL diversity: the 10 most-frequently posted URLs accounted for 41\% of all comments, compared to 80\% for greedy search.

Latency is another important aspect when choosing a decoding strategy for deployment. While beam search increased the posting frequency and diversity, inference became noticeably slower (median latency of 2 seconds). Given that this latency is prohibitive for interactive use in the IDE, we ultimately decided to use beam search for the code review system and greedy search for the IDE.

\subsection{Suppressing Outdated Best Practices}

After launching \toolname to around 3 thousand voluntary first-adopters, we noticed a large number of issues being filed by users within a few days. Many of these corresponded to a single URL\footnote{https://github.com/google/styleguide/blob/gh-pages/pyguide.md\#22-imports}, which describes best practices related to Python imports. However, the canonical source for some type names had changed in Python 3.9, and the best practice had also changed in early 2022. Since our training data stretches before 2022, it contained a number of best practice comments that were no longer applicable. We realized that this is a recurring pattern: as languages evolve, or new libraries are introduced, best practices evolve as well. One way of mitigating the problem is to filter out such data (whenever a rule changes), and retrain the model. This is however time and resource-consuming: it requires full data regeneration, model training, evaluation, and rollout. In the meantime, the ``outdated'' model needs to be either switched off, causing downtime of the system, or affected predictions need to be suppressed. Otherwise, the system could quickly lose developer trust. We opted for suppression of specific best-practice predictions, using conditional filtering (matching regular expressions on the source code) for two reasons. First, it can be dynamically deployed and immediately applied. Second, it allows for granular filtering of predictions.

\subsection{Independent Rating of Selected Comments}

After several months of early usage, we observed that the useful ratio plateaued at around 54\%. To understand the reasons, identify areas for improvement, and prepare for a wider deployment, we conducted an independent human rating study in April 2023, analyzing a sample of around 370 posted comments that received developer feedback during our first-adopters deployment.

To gather diverse perspectives on the usefulness of comments we recruited 15 raters---developers from partner teams. We asked them to rate \toolname's comments that received explicit user feedback. We did not show the original user feedback to the rater, to avoid biasing their evaluation.
The raters assessed each comment's usefulness based on the linked best practice and the surrounding code. We instructed them to focus on comment correctness, but also whether the comment would be actionable to them as an author (e.g., would they resolve a comment that is technically correct but does not seem worth resolving in a specific instance). They were encouraged to provide free-form feedback on each comment.

The useful ratio from the rater evaluation was 60\%, slightly higher than the 54\% from the developer feedback on the same comments, but well below our target of 80\% for wider deployment.

The most interesting finding from this study was that there were clear patterns of not useful comments. Here are some examples:

\smallskip
\textbf{Several topics or complex topic:} For example, one URL points to a section that describes multiple guidelines for interacting with the Python linter, including cases where it often triggers and ways to suppress it. An author may struggle to understand what specific guideline a posted comment is referring to and how to resolve it.
Similarly, the guidance on writing good function documentation in C++ is a full page of dense text. Raters frequently noted a disconnect between a best practice (and \toolname's concise summary) and the actual code, even when it contained a relevant violation.

\smallskip
\textbf{Importance of high-quality summaries:} Raters often found that \toolname's summary, which was generated by scraping the document source and sometimes missing, failed to adequately explain the relevance of the cited guideline to the comment/code. 

\smallskip
\textbf{Subjective and potentially contentious topic:} One example is avoiding flags in library code. Flags can cause problems when used in libraries, but some libraries are specifically designed to have many features configurable via flags. Additionally, legacy code may not adhere to this guideline and reviewers will not enforce it. The model did not learn these nuances and sometimes predicted a violation when an author added a new flag to an existing library.

\smallskip
\textbf{Systematic model error for some guidelines:} One interesting example is a guideline that promotes the use of the member function \textit{push\_back} over \textit{emplace\_back} for C++ vectors when both functions can be used with the same arguments to achieve the same effect. The model had learned to predict this, but it would also predict it in cases where \textit{emplace\_back} is warranted, and also when an unrelated type had a member function called \textit{push\_back}.

\smallskip
\textbf{Correct but low-value comments:} A missing period at the end of a sentence in a code comment is often allowed by human reviewers. While technically correct, asking the author to go back to their IDE and fix the issue may provide net negative value.

\medskip\noindent
The insights from the rater study informed two changes to \toolname. First, the rater study identified 17 non-actionable URLs, whose suppression increased the historical useful ratio from 54\% to 66\% on developer feedback, and from 60\% to 74\% on rater feedback. We further analyzed comments linked to similar, unrated URLs and suppressed an additional 5. Second, we reviewed and manually updated summaries for all frequently posted URLs. Together, these changes were sufficient to reach our target useful ratio of 80\% for the next stage of deployment.

\subsection{A/B Experiment}

In July 2023, we deployed \toolname to about half of all developers in the context of an A/B experiment. We randomly assigned developers to a an experiment group (\toolname enabled) and a control group (\toolname disabled). We randomized based on the last few digits of the SHA256 hash of the developers email address, and we verified that both groups did not differ in size and composition, including distribution of tenure, seniority, programming languages and business units. We also confirmed that none of the variables measured during the experiment differed between the control and the experiment group before the experiment began. The comment posting frequency during the experiment was in line with expectations (\cref{sec:initial_threshold_and_decoding_strategy_selection}).

We did not detect any statistically significant change in any of the following: total duration of code reviews, time developers actively spent on the code review, the number of comment-response iterations between the author and the reviewer. We did, however, detect a slight improvement in coding speed. We conjecture that the reduction in context switches to documentation leads to this positive effect. We leave a deeper investigation for future work.
Based on the results, we concluded that there are no adverse effects, and deployed \toolname to all developers in October 2023.

\begin{figure}
    \centering
    \includegraphics[width=\linewidth]{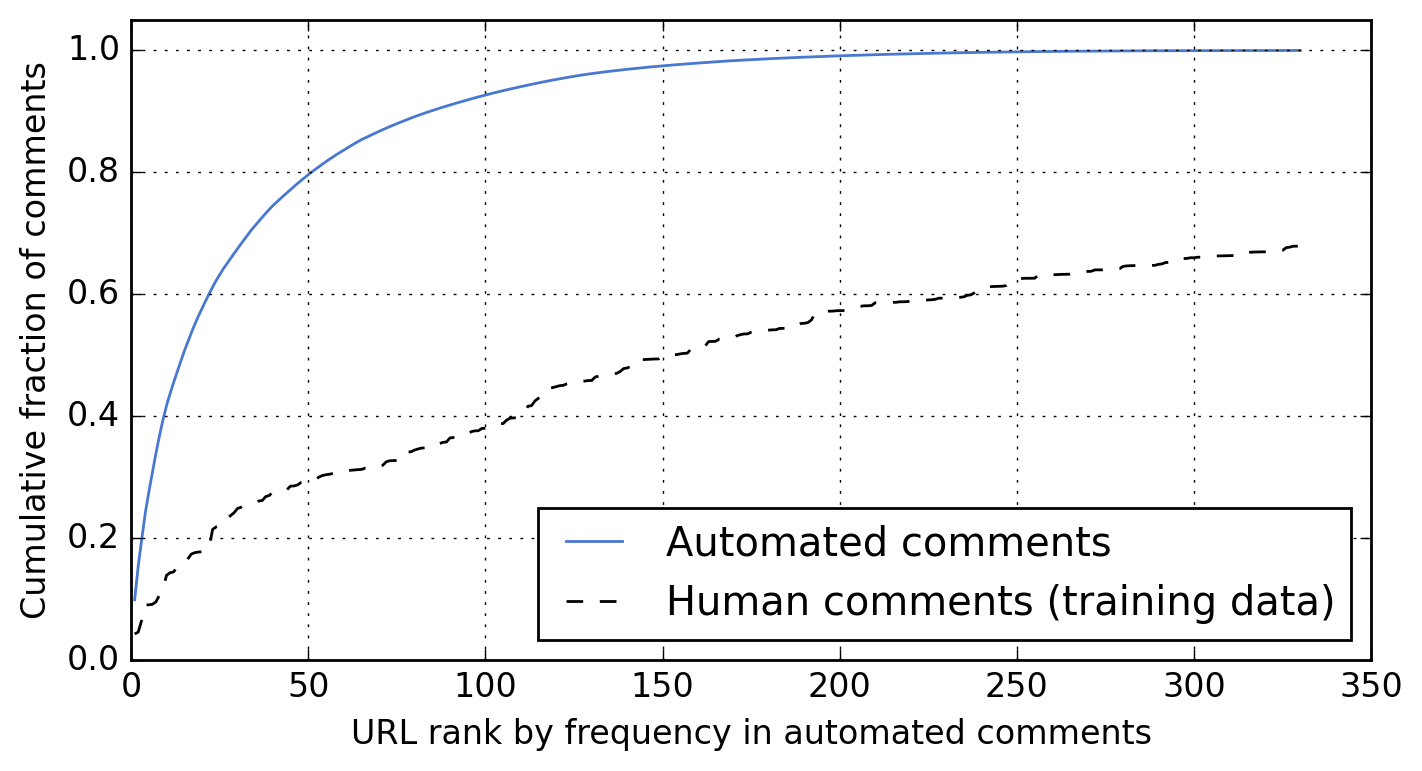}
    \caption{Cumulative distribution of comments per URL for the automated comments generated by \toolname in production and human comments in the training data.}
    \label{fig:tip_frequencies}
\end{figure}\section{Evaluation}
\label{sec:eval}

Based on the useful ratio and user feedback gathered since March 2023, we conclude that developers are generally satisfied with the comments produced by \toolname. We continuously refine our dataset preparation, thresholds, URL suppression and summarization by analyzing user feedback, to ensure that \toolname delivers high positive impact on the developer workflow.

Beyond developer satisfaction, with several months in wide release to all of \companyname, we evaluated three additional aspects of \toolname's performance:

\begin{enumerate}
    \item \textbf{Comment resolution:} How often do developers modify their code to resolve \toolname's posted comments?
    \item \textbf{\toolname vs. human comments:} How well do \toolname's comments cover the best practice documents that human reviewers reference in their comments?
    \item \textbf{\toolname vs. linters:} To what extent does \toolname's output go beyond the capabilities of traditional static analysis tools?
\end{enumerate}

\subsection{Comment Resolution}

Developers rarely give \textit{explicit feedback} on \toolname's comments by clicking the thumbs up/thumbs down buttons in the code review system and IDE, and the ``Please fix'' button in the code review system~(\cref{fig:example_critique_comment}): about 10\% of automated comments in the code review system and 2\% of diagnostics in the IDE received explicit feedback, which is comparable to other automated analyses at \companyname. At the same time, developers hover over approximately 50\% of the \toolname's IDE diagnostics, and prior work showed that developers often resolve automated comments without explicit feedback~\cite{petrovic2023resolution}.
To assess how often developers resolve \toolname's comments, we conducted an offline analysis, estimating the ratio of comments resolved by subsequent code changes.

To analyze comment resolution, we extracted historical changes focused on files with automated comments from \toolname. For each, we extracted the initial snapshot where the comment was posted and the snapshot that the developer eventually merged into the codebase. Each comment spans a specific range of lines. We used an automated AST-based line mapping approach~\cite{petrovic2023resolution} between these snapshots, to identify comments that the model originally predicted on the first snapshot, but did not predict on the merged snapshot. Such pairs of snapshots indicated that a comment may have been resolved, but it is possible that unrelated code changes could have led to a specific comment no longer being predicted.

An automated analysis of 6000 snapshot pairs revealed that in 50\% of cases the comment was absent from the submitted snapshot on the lines it was originally posted. We manually inspected a random sample of 40 such pairs. We found that in 80\% of cases, a change made by the author directly resolved the issue described by the posted comment. Therefore, we estimate that the comment-resolution rate is about 40\%, which is significantly larger than the ratio of comments with explicit positive feedback to all comments.

\subsection{\toolname vs. Human Comments}

\Cref{fig:tip_frequencies} compares the cumulative distribution of comments (per unique URL to a best practice document) for the automated comments generated by \toolname in production and human comments in the training data. The x-axis is the rank of the URL when all URLs ever used in automated comments are sorted by frequency. For example, the most frequently used URL has rank 1, and it accounts for 9.9\% of all automated comments. This same URL appeared in 4.3\% of the human created comments in the training data. In total, \toolname has created comments for 330 distinct URLs. The set of URLs used by \toolname covers 68\% of historical human comments with a best practice URL. This is a good result: it demonstrates that \toolname is not focusing on obscure best practices that are rarely referenced by reviewers.

On the other hand, despite utilizing beam search, URL diversity remains relatively low. The top-85 URLs make up 90\% of comments created by \toolname. The same set of URLs cover 35\% of human comments with best practice URLs. Improving URL diversity and coverage of best practices in automated comments while maintaining accuracy and low latency is one of our top priorities.

\subsection{\toolname vs. Linters}
\begin{figure}
    \centering
    \includegraphics[width=\linewidth,trim={18pt 0 0 0},clip]{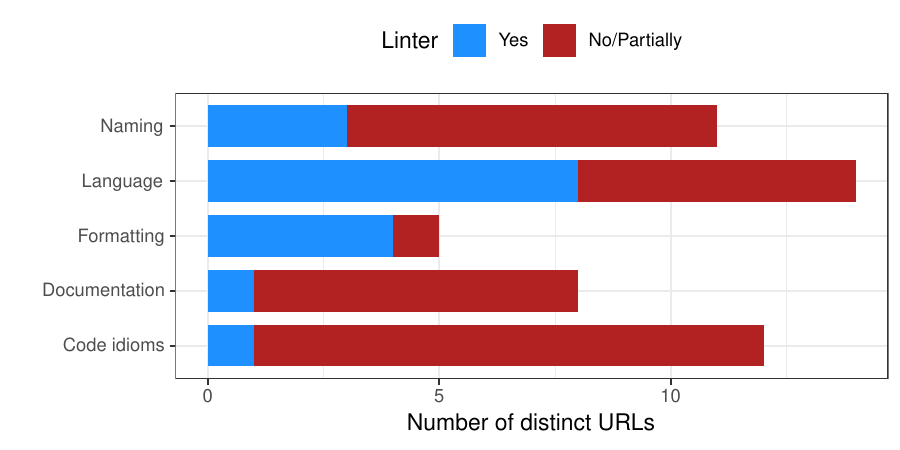}
    \caption{The top-50 most frequently predicted URLs categorized into types. Linter indicates whether a linter that detects a violation exists or can easily be built.}
    \label{fig:tip_types}
\end{figure}
To understand to what extent \toolname provides value beyond linters that can efficiently and precisely check some of the best practices, we sampled the top-50 most frequently predicted violations---that is, the top-50 URLs in~\cref{fig:tip_frequencies}. For each sampled URL, we inspected its best practice document and determined (1) the best-practice type (\cref{sec:intro}) and (2) whether a linter that detects a corresponding violation exists or can be easily built. Specifically, three authors, each with over 10 years of experience in building static analysis tools, read the documentation and independently categorized the URLs. There were no disagreements on the best-practice type, but there were disagreements on whether a linter can be easily built for about 15\% of URLs. The three authors resolved these disagreements through majority vote and discussion. Disagreements stemmed from ambiguous best practices, and those with multiple guidelines. For example, while checking the presence of code documentation is relatively straightforward, reasoning about justified exceptions and clarity of content may not.

\Cref{fig:tip_types} shows the distribution of the 50 sampled URLs, broken down by type and whether violations can be detected by a linter. For 33/50 (66\%) of these best practices, violation detection is beyond the scope of traditional static analysis.\section{Lessons learned}
\label{sec:lessons}

Based on our experience developing and deploying \toolname, we summarize a few key lessons learned:

\begin{itemize}

\item \textbf{Complementing traditional analyses}: \toolname's LLM-backed approach generates comments for 68\% of best practices frequently referenced by human reviewers. Many of these are out of scope for traditional static analyses.

\item \textbf{Intrinsic evaluation vs. real-world performance}: 
Intrinsic evaluations and real-world performance can diverge significantly:
our intrinsic evaluation, using a dataset of real-world human comments together with a state of the art model architecture and training process, indicated a promising model, but our extrinsic evaluations and system improvements proved essential for a successful deployment.

\item \textbf{Monitoring user acceptance is critical}: Even a few negative user experiences can erode trust in an automated system. Continuously monitoring and analyzing real-world feedback was crucial in detecting such instances and identifying remedies. In the case of \toolname, a simple suppression mechanism was sufficient to strongly improve user acceptance to over 80\% without major sacrifices in efficacy.

\end{itemize}\section{Related Work}
\label{sec:related}

Johnson~\cite{johnson1977lint} introduced the C linter almost 50 years ago in 1977. In those 50 years, a considerable body of research on automated static analysis was produced: a recent literature review by Heckman and Williams~\cite{HECKMAN2011363} identified 17,571 papers. Many studies explore how developers interact with static analysis. \citet{johnson2013don} explore challenges developers face when trying to use static analysis. The results of their study highlights the importance of good integration into existing developer workflows and the importance of developing and maintaining trust in the tool. \citet{vassallo2020developers} explore how developers interact with static analysis in different contexts, including coding and code review. They too find that integration into existing workflows plays a major role in developers willingness to use the tools and that high quality of results is extremely important. \citet{beller2016analyzing} studied usage of static code analysis in a large number of open source projects. Among other findings, they highlight that how automated analysis is and should be used varies based on the programming language.

In contrast, using machine learning for code analysis is a comparatively new and less understood field. A number of recent publications
(e.g.,
\citet{hong2022commentfinder},
\citet{li2022auger},
\citet{10.1145/3540250.3549081},
\citet{thongtanunam2022autotransform},
\citet{tufano2024code}, and
\citet{tufano2022using})
report on model evaluations and propose tools for automated code review.
While these models and the review comment generation task are very similar to the model presented in this paper, evaluations largely focused on historical datasets. As discussed in \cref{sec:evaluation_historical_comments} an intrinsic evaluation on only historical comments is somewhat limited and can sometimes fail to predict real-world performance. Another recent publication by~\citet{C2CGoogle2024} presents an evaluation of a live system, but for the opposite task: creating code from comments rather than comments from code.\section{Conclusion}
\label{sec:discussion}
\label{sec:conclusion}

Verifying that code adheres to best practices is a common task in modern code review processes. While some best practices can be automatically verified with traditional tools such as linters, many require the knowledge and judgement of experienced developers, which requires time and effort.

This paper reports on our experience developing, deploying, and evaluating \toolname, an LLM-backed code review assistant system. Specifically, it lays out the entire process from task and model design, over intrinsic evaluations and system calibrations, to a staged roll out and end-user evaluation.

The evaluation results show that it is feasible to develop an end-to-end system with capabilities well beyond traditional tools while achieving a high degree of end-user acceptance. These results are a promising first step towards the deployment of sophisticated code-review assistants and automated code reviews.

Our priority was to ensure a positive developer experience by designing \toolname to have very high precision. While recall was not the primary focus, we recognize its significance and plan to explore what changes in the model and system architecture can improve recall.
For example, the model we used in 2022 was state of the art at the time. However, it has a limited context window of 2048 tokens which suffices for only around 200 lines of code. Current state of the art models have context windows of tens of thousands of tokens during training and over a million tokens during inference. This leap opens up opportunities for new features and significant improvement in existing ones.\section{Acknowledgements}
\label{sec:ack}

This work is the result of years of collaboration between teams in Google Core Systems and Google DeepMind. 
We are grateful for the support and advice of all our team members and leadership, including Alberto Elizondo, Alexander Frömmgen, Ballie Sandhu, Chandu Thekkath, Chris Gorgolewski, David Tattersall, Ilya Cherny, Jacob Austin, Katja Grünwedel, Kristóf Molnár, Lera Kharatyan, Luka Rimanić, Madhura Dudhgaonkar, Marc Brockschmidt, Marcus Revaj, Maxim Tabachnyk, Nina Chen, Niranjan Tulpule, Nitya Ramani, Paige Bailey, Pavel Sychev, Pierre-Antoine Manzagol, Quinn Madison, Roger Fleig, Satish Chandra, Savinee Dancs, Stoyan Nikolov, Subhodeep Moitra, and Vaibhav Tulsyan.
\balance
\bibliographystyle{ACM-Reference-Format}
\bibliography{venues,references_arxiv}

\end{document}